\documentclass[cernpreprint,backref=false,texlive=2017,american,texmf]{atlasdoc}
 
\usepackage[subcaption]{atlaspackage}
\usepackage{atlasbiblatex}
 
\usepackage{atlasphysics}
 
\graphicspath{{logos/}{figures/}}
 
\usepackage{ANA-STDM-2018-16-PAPER-defs}

\AtlasTitle{Observation and measurement of forward proton scattering in association with lepton pairs produced via the photon fusion mechanism at ATLAS}

\AtlasAbstract{
The observation of forward proton scattering in association with lepton pairs
($e^+e^-+p$ or $\mu^+\mu^-+p$) produced via photon fusion is presented.
The scattered proton is detected by the ATLAS Forward Proton spectrometer
while the leptons are reconstructed by the central ATLAS detector.
Proton-proton collision data recorded in 2017 at a center-of-mass energy of $\sqrt{s} = 13$~TeV are analyzed,
corresponding to an integrated luminosity of 14.6~fb$^{-1}$.
A total of 57 (123) candidates in the $ee+p$ ($\mu\mu+p$) final state are selected,
allowing the background-only hypothesis to be rejected with a significance exceeding 5 standard deviations in each channel.
Proton-tagging techniques are introduced for cross-section measurements in the fiducial detector acceptance,
corresponding to
$\sigma_{ee+p} = 11.0\pm 2.6\left(\text{stat}\right) \pm 1.2\left(\text{syst}\right) \pm 0.3\left(\text{lumi}\right)$ and
$\sigma_{\mu\mu+p} = 7.2\pm 1.6\left(\text{stat}\right) \pm 0.9\left(\text{syst}\right) \pm 0.2\left(\text{lumi}\right)~\text{fb}$
in the dielectron and dimuon channel, respectively.
}
 
\author{The ATLAS Collaboration}
 
\AtlasRefCode{ANA-STDM-2018-16}
 
\PreprintIdNumber{CERN-EP-2020-167}

\arXivId{2009.14537}

\AtlasJournalRef{Phys. Rev. Lett. 125 (2020) 261801}
\AtlasDOI{10.1103/PhysRevLett.125.261801}

\hypersetup{pdftitle={ATLAS document},pdfauthor={The ATLAS Collaboration}}
 
\begin{document}
 
\maketitle

Electromagnetic fields sourced by protons at the Large Hadron Collider (LHC)
are sufficiently intense to exceed the Schwinger limit of $10^{18}$~V\,m$^{-1}$~\cite{BreitWheeler1934,HeisenbergEuler1936,SchwingerLimit}
and produce lepton pairs via photon fusion, $\gamma\gamma\to\ellell$, where $\ell$ denotes electrons or
muons~\cite{STDM-2014-13,STDM-2016-13,CMS-FWD-11-004,CMS-FWD-10-005}.
This process occurs in a wide range of astrophysical phenomena, such as cosmic gamma rays~\cite{Gould1967,dwek2013extragalactic}
and neutron stars~\cite{Ruffini:2009hg,Kaspi:2017fwg}.
Measurements of $\gamma\gamma\to\ellell$ at the LHC provide a unique laboratory probe of these natural phenomena and are fundamental tests of
quantum electrodynamics~\cite{PhysRevD.7.3485,Budnev:1974de,Piotrzkowski:2000rx,Khoze:2001xm,deFavereaudeJeneret:2009db,Aaij:2018arx}.
These complement lower-energy probes using
heavy-ion collisions~\cite{HION-2018-11,HION-2018-19,CMS-FSQ-16-012,Acharya:2019vlb,Baltz:2007kq,Afanasiev:2009hy,Adam:2019mby,Beresford:2019gww,Dyndal:2020yen}
and high-intensity laser
beams~\cite{Burke:1997ew,Ruf:2009zz,Altarelli:2019zea,Abramowicz:2019gvx}.
A hallmark prediction of photon fusion processes at the LHC is the forward scattering of incident protons.
Near-beam instruments known as proton spectrometers can detect the scattered protons, which is a technique referred to as proton tagging.
The CMS and TOTEM Collaborations reported proton-tagged dielectron (dimuon) production with $2.6\sigma(4.3\sigma)$ significance,
which exceeds $5\sigma$ when statistically combined~\cite{CMS-PPS-17-001}, but no cross sections were measured.
Previous measurements of $\gamma\gamma\to\ellell$ by the ATLAS Collaboration
were performed without proton tagging~\cite{STDM-2014-13,STDM-2016-13}.
 
Measuring proton-tagged dilepton production, $p p \to p (\gamma\gamma\to\ellell) p^{(*)}$,
where $p^{(*)}$ denotes a proton that remains intact or dissociates following electromagnetic excitation, is important for several reasons.
Predictions of photon fusion processes have significant uncertainties
associated with modeling strong-force interactions
between scattered protons, which suppress cross sections by factors known as
soft-survival probabilities~\cite{Khoze:2013dha,Harland-Lang:2016apc,Dyndal:2014yea,Harland-Lang:2015cta}.
This suppression is poorly constrained, especially at high $\gamma\gamma$ invariant masses important for new physics searches,
as existing probes \emph{indirectly} infer dissociation rates using only central-detector information~\cite{STDM-2014-13,STDM-2016-13,CMS-FWD-11-004,CMS-FWD-10-005}.
Proton tagging overcomes this longstanding experimental ambiguity by \emph{directly} detecting the scattered protons.
Detecting a proton also directly suppresses background processes and events involving proton dissociation,
while providing information on the initial $\gamma\gamma$ system independently of central-detector information.
The successful demonstration of proton-tagging techniques for cross-section measurements
accomplishes the crucial first step toward a diverse program using proton tagging in measurements of
Standard Model processes~\cite{Cox:2007sw,Albrow:2008pn,Trzebinski:2015joa,Tizchang:2020tqs,STDM-2015-10,CMS-FSQ-13-008} and
searches for new phenomena~\cite{Heinemeyer:2007tu,HarlandLang:2011ih,Baldenegro:2018hng,Beresford:2018pbt,Harland-Lang:2018hmi}.
 
This Letter introduces proton tagging for cross-section measurements of $p p \to p (\gamma\gamma\to\ellell) p^{(*)}$.
The ATLAS Forward Proton (AFP) spectrometer detects one of the intact protons
and the central ATLAS detector reconstructs the leptons.
The dataset was collected in 2017 and corresponds to 14.6~fb$^{-1}$ of $\sqrt{s}= 13~\TeV$ proton-proton $(pp)$ collisions.
The average number of interactions per bunch crossing was 36.
Several methods specific to proton tagging are introduced:
in situ calibration of proton kinematics using the dimuon system,
a novel data-mixing background estimation method,
and tag-and-probe determination of the AFP reconstruction efficiency.
 
The ATLAS experiment~\cite{PERF-2007-01,ATLAS-TDR-2010-19,PIX-2018-001} is a general-purpose particle detector
with nearly $4\pi$ coverage~\cite{atlas-footnote} around the interaction point.
It comprises an inner detector tracker, calorimeters, and a muon spectrometer.
A two-level trigger system~\cite{TRIG-2016-01} is employed to select events
containing same-flavor lepton pairs,
each lepton with $\pt^{e\left(\mu\right)} > 17\left(14\right)~\GeV$~\cite{ATL-DAQ-PUB-2018-002,TRIG-2018-01,TRIG-2018-05},
after which standard data-quality requirements are applied~\cite{DAPR-2018-01}.
 
The AFP spectrometer~\cite{Adamczyk:2017378,ATL-PHYS-PUB-2017-012}
consists of four tracking units located along the beam pipe at $z = \pm 205$~m and $\pm 217$~m,
referred to as near and far stations, respectively. The $+z\left(-z\right)$ direction is labeled side $A$ ($C$).
Each station houses a silicon tracker comprising four planes of edgeless silicon pixel sensors~\cite{Lange:2015xla,Lange:2016dro,GarciaSciveres:2011zz,Zivkovic_2012}.
The sensors have $336 \times 80$ pixels with area $50 \times 250~\mu$m$^2$.
The direction normal to each sensor is tilted $14^\circ$
relative to the beam to improve hit efficiency and $x$-position resolution,
resulting in an overall spatial resolution of $\sigma_x = 6~\mu$m~\cite{Grinstein:2016hmq}.
Movable near-beam devices at each station, known as Roman pots, insert the tracker along the $x$ direction in the beampipe.
Data taking with the AFP commences once the trackers are at a position
where the innermost silicon edge is within 2~mm of the beam center during stable beams.
Data quality for this analysis requires that every AFP station has at least three silicon planes operational at high voltage,
and the AFP data acquisition system~\cite{Kocian:2017ygu} must report no problems.
 
Simulated events of the exclusive signal $pp \to p(\gamma\gamma\to\ellell)p$
were produced using the \textsc{herwig{\scriptsize7}} Monte Carlo (MC) generator~\cite{Bahr:2008pv,Bellm:2015jjp}.
The single-dissociative signal $pp \to p(\gamma\gamma\to\ellell)p^{*}$ was generated using \textsc{lpair{\scriptsize4.0}}~\cite{Vermaseren:1982cz},
with proton dissociation modeled using the Brasse et al.~\cite{BRASSE1976413} and Suri-Yennie~\cite{SURI1972243} structure functions
interfaced with \textsc{jetset{\scriptsize7.408}}~\cite{SJOSTRAND199474,ANDERSSON198331}.
Simulation of these processes is detailed in Ref.~\cite{STDM-2016-13}.
To model the central-detector response, the exclusive signal sample underwent full detector simulation based on \textsc{geant}{\scriptsize4}~\cite{Agostinelli:2002hh}.
The single-dissociative samples employed a fast simulation~\cite{SOFT-2010-01}, which uses a parametrization of the calorimeter response~\cite{ATL-PHYS-PUB-2010-013}.
The response of the AFP spectrometer is modeled by a fast simulation, where a Gaussian smearing
is applied to track positions based on the AFP spatial resolution.
Simulated samples include the effect on the central detector of
multiple $pp$ interactions in the same and neighboring bunch crossing (pileup), as detailed in Ref.~\cite{STDM-2016-13}.
 
Reconstructed events must contain at least one interaction vertex with two or more associated inner-detector tracks
that satisfy $\pt > 500~\MeV$, $|\eta|<2.5$, and the ``Loose'' criterion~\cite{ATL-PHYS-PUB-2015-051,PERF-2015-08}.
Electrons (muons) must satisfy $\pt > 18\left(15\right)~\GeV$, $|\eta|<2.47\left(2.4\right)$,
the ``LooseAndBLayer''~\cite{PERF-2017-01} (``Medium''~\cite{PERF-2015-10})
identification criterion, and $|z_0\sin\theta| < 0.5$~mm~\cite{PV-footnote}.
Electrons sharing an inner-detector track with a muon are discarded.
To suppress fake and/or nonprompt lepton backgrounds,
remaining electrons (muons) must satisfy transverse impact parameter significance $\left|d_0/\sigma_{d_0}\right| < 5\left(3\right)$
and isolation requirements described in Ref.~\cite{SUSY-2018-16} (Ref.~\cite{SUSY-2018-32}).
Electrons must also satisfy ``Medium'' identification~\cite{PERF-2017-01}.
Small corrections are applied to leptons in simulated samples
to match reconstruction and trigger efficiencies measured in data, as described in Refs.~\cite{PERF-2017-01,PERF-2015-10}.
 
Selected events must have exactly two same-flavor leptons with opposite electric charge ($e^+e^-$ or $\mu^+\mu^-$)
and be matched to the leptons that triggered the event.
To suppress quarkonia and $Z$ boson resonances,
the dilepton invariant mass must satisfy $\mll > 20$ and $\mll \notin [70, 105]~\GeV$.
To select events compatible with $pp \to p(\gamma\gamma\to \ellell)p^{(*)}$ processes based on the simulated signals,
the dilepton transverse momentum must satisfy $\ptll < 5~\GeV$.
This set of criteria is referred to as the preselection.
Signal event candidates must additionally have small acoplanarity $\acoll = 1-|\Delta \phi_{\ell\ell}| /\pi < 0.01$.
These events must have no inner-detector tracks ($\Ntracks = 0$) that satisfy $\Delta R(\text{track}, \ell) > 0.01$ for both leptons
and $|z_0^\text{track} - z_0^{\ell\ell} | < 0.5$~mm,
where $z_0^\text{track}$ is the track $z_0$ position
and $z_0^{\ell\ell} = ( z_0^{\ell_1} + z_0^{\ell_2} ) / 2$ with $\ell_{1, 2}$ denoting the two leptons.
The expected proton energy loss based on lepton kinematics $\xill$ is determined
from $\mll$ and the dilepton rapidity $\yll$ by momentum conservation $\xill^{\pm} = ( \mll / \sqrt{s} ) e^{ \pm \yll }$,
where $+$ ($-$) corresponds to the proton on side $A$ ($C$).
 
Reconstruction of scattered protons combines information from the AFP tracker and LHC magnet lattice~\cite{Evans:2008zzb}.
Protons transported to the AFP leave hits
in the silicon tracker, which are processed by clustering and track-finding algorithms detailed in Ref.~\cite{Lange:2016dro}.
Tracks are reconstructed from clusters in at least two planes.
Small corrections of around 0.1~mm are applied to ensure the cluster positions between planes are compatible within the spatial resolution.
The proton transport function $x_\text{AFP} = T(\xiAFP)$ relates the track $x$ position $x_\text{AFP}$
to the fractional energy loss of the scattered proton $\xiAFP = 1-E_\text{scattered} / E_\text{beam}$,
where $E_\text{scattered}$ ($E_\text{beam}$) is the scattered (beam) proton energy.
The LHC magnets and beam optics~\cite{LHCOpticsWeb} govern the form of $T(\xiAFP)$~\cite{transport-footnote},
which is simulated in the \textsc{mad-x} package~\cite{Herr:744163,MADXmanual} with further details discussed in
Refs.~\cite{Adamczyk:2017378,Staszewski:2009sw,Staszewski:2014hua}.
Determination of $\xiAFP$ uses both the near and far stations if tracks are within their common acceptance,
otherwise only the far station is used.
 
The absolute scale of $E_\text{scattered}$ depends on the closest separation $x_0^s$ between each AFP station $s$ and the beam center~\cite{Staszewski:2014hua}.
The beam positions relative to the detectors were determined in dedicated runs with beam-based alignment procedures~\cite{PhysRevSTAB.15.051002}
using beam loss monitors~\cite{Zamantzas2007}, and cross-checked with beam position monitor measurements~\cite{PhysRevAccelBeams.20.081002}.
There were three data-taking periods in 2017.
In the first data-taking period, the $x_0^s$ values were initially set to
$-4.0\left(-3.0\right)$~mm on side $A$ and $-3.8\left(-2.9\right)$~mm on side $C$ for the near (far) stations;
during a second data-taking period, all stations were moved 0.5~mm closer to the beam to improve acceptance.
This first (second) data-taking period corresponds to 5\% (17\%) of the analyzed dataset.
For the remaining dataset, the far stations were moved a further 0.2~mm toward the beam.
The initially measured $x_\text{AFP}$ values relative to $x_0^s$ are calibrated in situ using the dimuon data sample passing the signal event selection.
The $x_{\ell\ell}^s - x_\text{AFP}^s$ distribution is peaked for signal processes due to the kinematic correlation between $x_{\ell\ell}^s$ and $x_\text{AFP}^s$,
where $x_{\ell\ell} = T(\xill)$ is the expected position calculated using the transport function.
Additive corrections are applied to $x_\text{AFP}^s$ in data to center the maximum of the peak at zero.
These corrections are found to be $-0.28\left(-0.34\right)$~mm on side $A$ and
$-0.17\left(-0.36\right)$~mm on side $C$ for the near (far) stations.
Selected dielectron events are used to verify that the signal is centered at zero.
After applying these corrections, the lower value of the acceptance corresponds to $\xiAFPA > 0.028\left(0.018\right)$ on side $A$
and $\xiAFPC > 0.026\left(0.019\right)$ on side $C$ for the near (far) stations.
The upper value of the acceptance is bounded by $\xiAFP < 0.12$ due to the presence of beam collimators~\cite{Adamczyk:2017378}.
 
To select events with one or more proton candidates,
the $\xill$ and $\xiAFP$ values for at least one AFP side are required to be within the range $[0.02, 0.12]$.
If there is more than one proton candidate on the same AFP side,
which occurs in 35\% of selected events, the proton with $\xiAFP$ closest to $\xill$ is chosen.
Proton-tagged dilepton candidates, denoted $\ell\ell + p$,
are selected by requiring kinematic matching on at least one AFP side, $\left | \xiAFP - \xill \right| < 0.005$,
which retains (rejects) more than 95\% (85\%) of the signal (background).
 
The dominant source of background after this selection
arises from lepton pairs produced in a $pp$ interaction different from that of the detected proton.
In this case, the lepton pairs are produced via the Drell-Yan mechanism
as well as $\gamma\gamma\to\ellell$ processes, in which
any outgoing protons are either outside the AFP acceptance or not reconstructed in AFP due to detector inefficiency.
These events are collectively referred to as combinatorial backgrounds and are estimated using a data-driven method.
A mixed-data sample is constructed by randomly pairing each measured $\xill$ value, passing
AFP acceptance $\xiAFP \in [0.02, 0.12]$, with 100 values of $\xiAFP$ from a large control sample of $>10^6$ events.
This control sample is constructed from the preselected events and requiring $\acoll > 0.01$.
The 123 selected data events failing kinematic matching, $|\xiAFP - \xill| > 0.005$, result mostly from combinatorial background processes,
which are used to normalize the mixed-data sample using a background-only profile-likelihood fit~\cite{Cowan:2010js,Baak:2014wma}.
 
Systematic uncertainties in the background normalization arise from the limited size of the data sample satisfying $|\xiAFP - \xill| > 0.005$.
An uncertainty in the background shape arises from kinematic changes in the control sample of protons due to the acoplanarity requirement.
This uncertainty is estimated by replacing the $\acoll > 0.01$ condition with $\Ntracks \geq 1$
and comparing the two background predictions in the region $|\xiAFP - \xill| < 0.005$; they are found to differ by 14\%.
Further shape uncertainties arise from instrumental effects,
which are expected to be dominated by the sensitivity to the number of interactions per bunch crossing, $\mu$.
The background predictions for $\mu < 35$ and $\mu \geq 35$ are found to differ by 8\% in the $|\xiAFP - \xill| < 0.005$ region.
These two shape differences are assigned as additional uncertainties.
 
The background estimation method is validated by applying it to the orthogonal $\mll~\in~[70, 105]~\GeV$ region.
The region $|\xiAFP - \xill| > 0.005$ is dominated by Drell-Yan events, which have no correlated protons.
In this region, the data and prediction from the mixed-data sample are found to be
compatible within the uncertainties across the $\xiAFP-\xill$ range for both sides $A$ and $C$.
 
\begin{figure}
\centering
\includegraphics[width=0.68\textwidth]{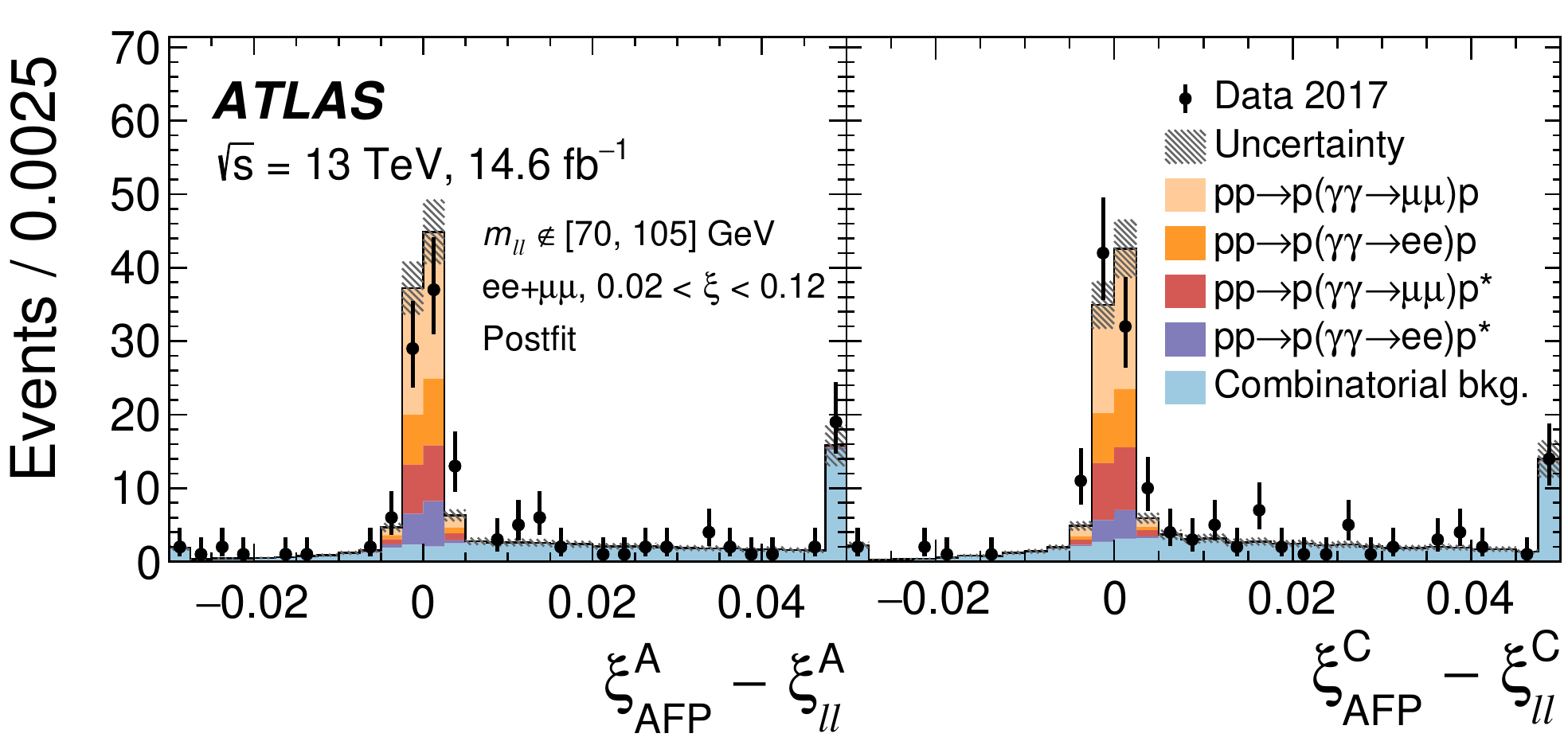}
\caption{
\label{fig:paper_fig_1}
Distributions of $\xiAFP - \xill$ with $\xill$ and $\xiAFP$ satisfying $[0.02, 0.12]$
for side $A$ (left) and side $C$ (right).
The total prediction comprises the signal and combinatorial background
processes, where $p^{*}$ denotes a dissociated proton.
The simulated predictions are normalized to data to illustrate the expected signal composition.
The first (last) bin includes underflow (overflow).
The hatched band indicates the combined statistical and systematic uncertainties of the prediction.
Error bars denote statistical uncertainties of the data.
}
\end{figure}

\begin{figure}
\centering
\includegraphics[width=0.58\textwidth]{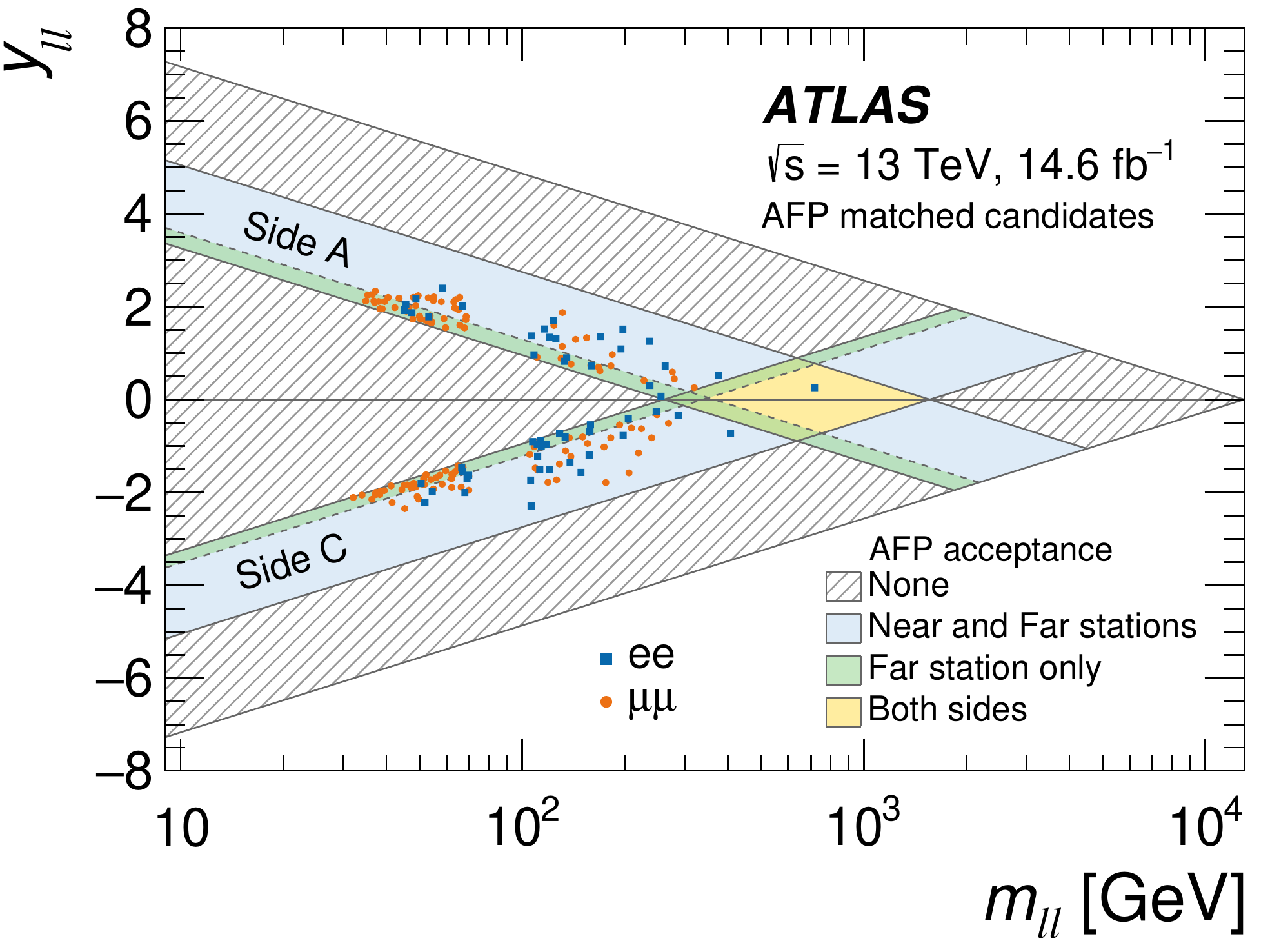}
\caption{The 57 (123) $ee$ ($\mu\mu$) data event candidates in the dilepton rapidity $\yll$ vs $\mll$ plane
satisfying event selection and kinematic matching, $|\xiAFP -\xill| < 0.005$, on at least one side.
Shaded (hatched) areas denote the acceptance (no acceptance) for the AFP stations indicated in the legend.
Areas neither shaded nor hatched correspond to $\xi \notin [0, 1]$.
}
\label{fig:paper_fig_2}
\end{figure}

\begin{figure}
\centering
\includegraphics[width=0.3\textwidth]{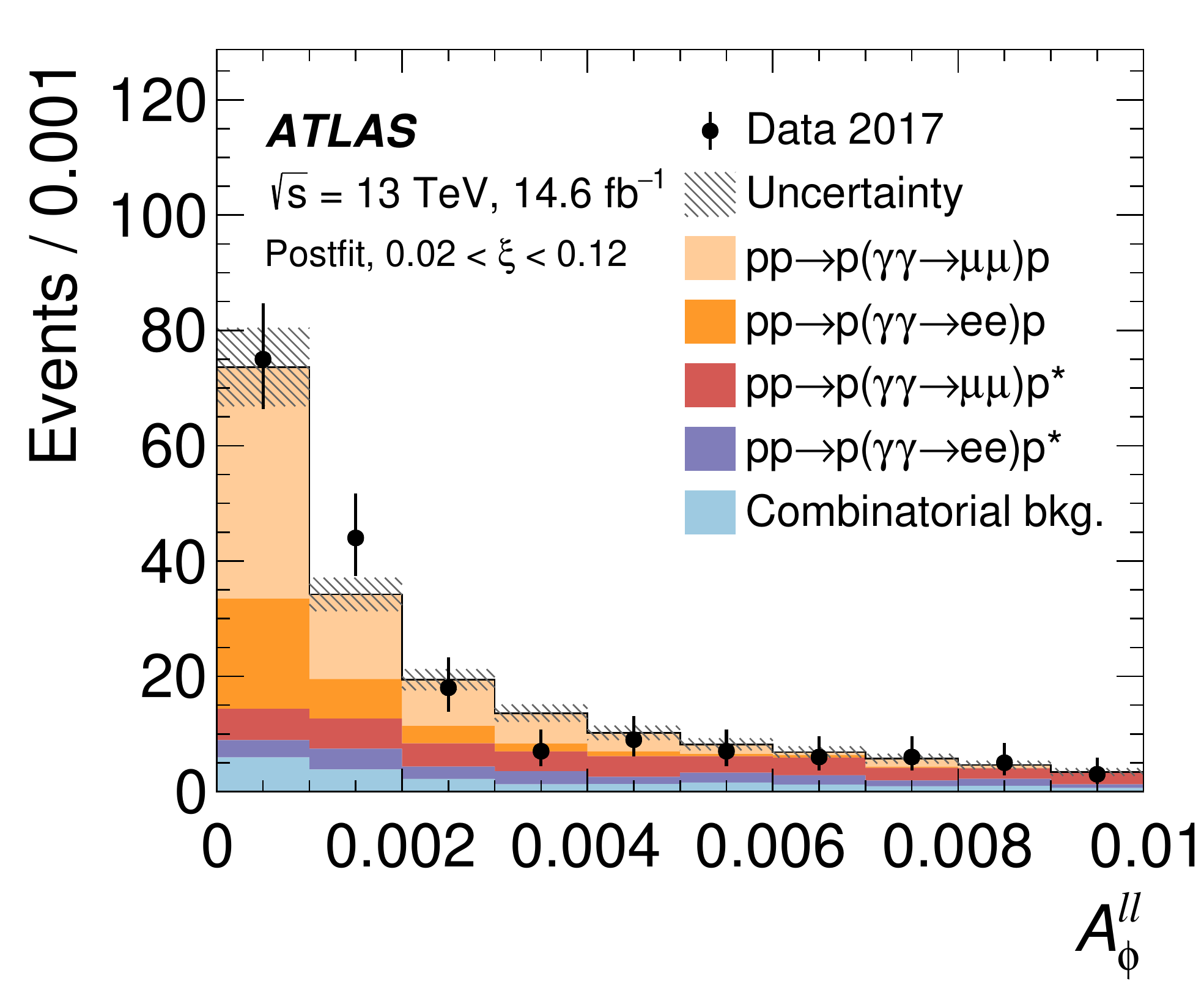}
\includegraphics[width=0.3\textwidth]{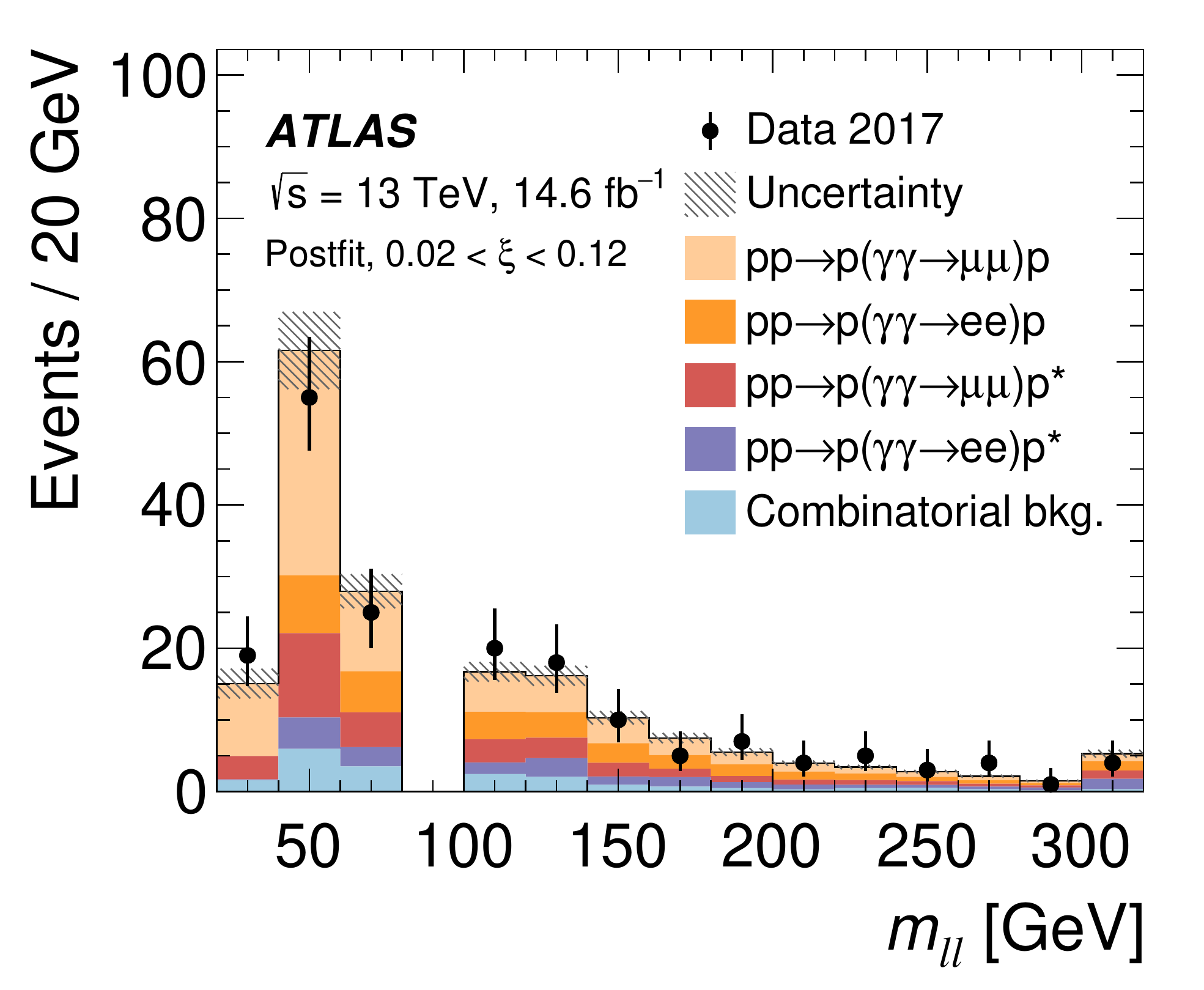}
\includegraphics[width=0.3\textwidth]{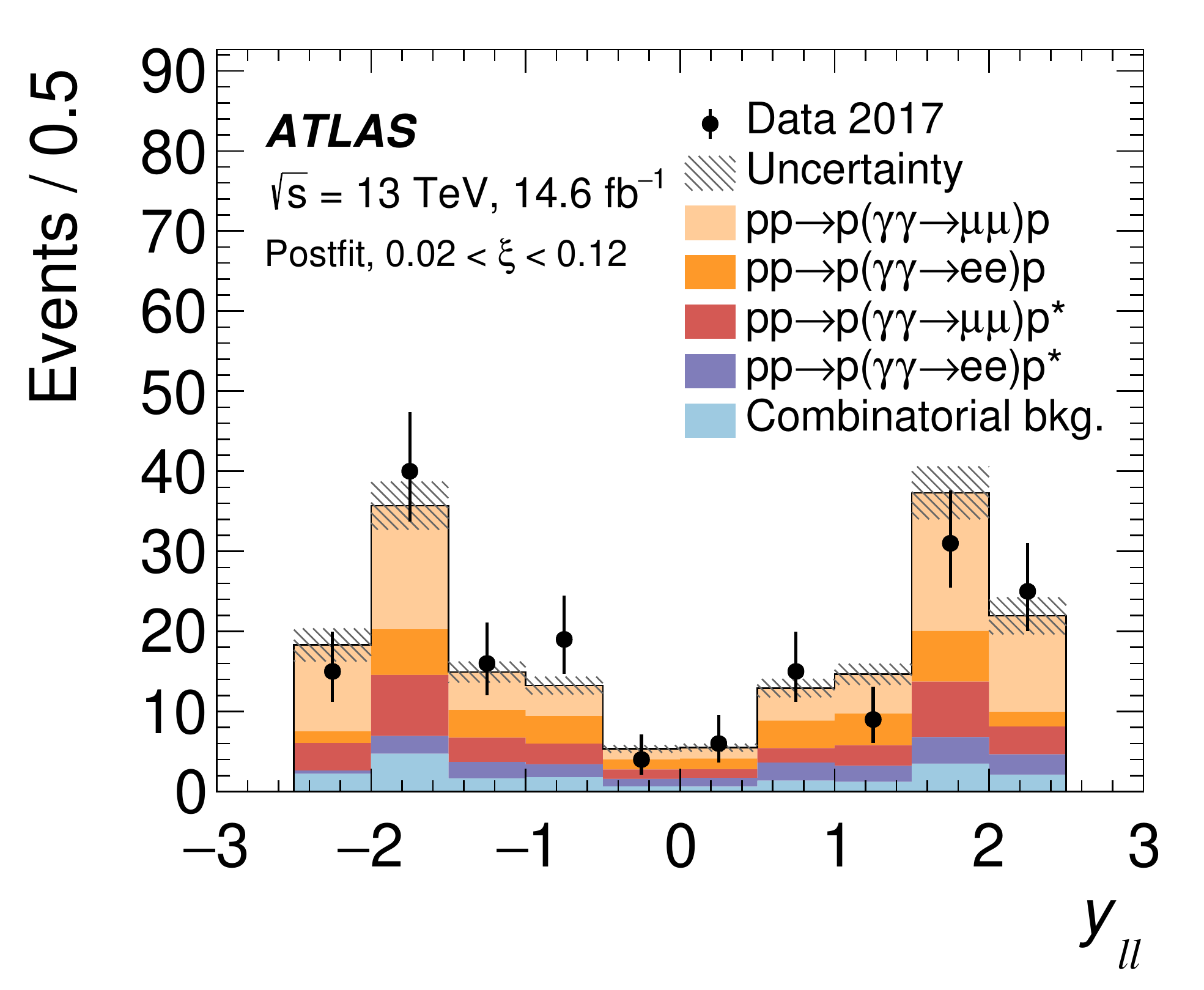}
\caption{
\label{fig:paper_fig_3}
Distributions of dilepton acoplanarity $\acoll$ (left), invariant mass $\mll$ (center), rapidity $\yll$ (right)
satisfying $\xill, \xiAFP \in [0.02, 0.12]$, and $|\xiAFP - \xill|<0.005$ for at least one AFP side.
Events with $70 < \mll < 105~\GeV$ are vetoed.
The total prediction comprises the signal and combinatorial background
processes, where $p^{*}$ denotes a dissociated proton.
The simulated predictions are normalized to data to illustrate the expected signal composition.
The rightmost bin of the $\mll$ distribution includes overflow.
The hatched band indicates the combined statistical and systematic uncertainties of the prediction.
Error bars denote statistical uncertainties of the data.
}
\end{figure}

After applying the event selection including kinematic matching, $|\xiAFP - \xill| < 0.005$,
a total of \NobsEE~(\NobsMM) candidates in the $ee+p$ ($\mu\mu+p$) final state are observed
compared with a background-only expectation of $\NbkgEE$~$(\NbkgMM)$ events.
Using the asymptotic profile-likelihood method~\cite{Cowan:2010js,Baak:2014wma},
the background-only hypothesis is rejected with a significance exceeding $5\sigma$ in each channel~\cite{signif-footnote}.
This provides direct evidence of forward proton scattering in association with electron and muon pairs produced via photon fusion.
The $\xiAFP - \xill$ distributions of data, signal, and background at detector level before kinematic matching are shown in Fig.~\ref{fig:paper_fig_1}.
To illustrate the expected composition of the signal, the simulated samples are
normalized to data with sides $A$ and $C$ combined and fit separately in the $ee$ and $\mu\mu$ channels.
Figure~\ref{fig:paper_fig_2} displays positions in the $\yll$--$\mll$ plane of data candidates satisfying $|\xiAFP - \xill| < 0.005$ on at least one side
and the corresponding acceptance regions of the four AFP stations.
The highest-mass $ee$ candidate has an invariant mass $\mll = 717~\GeV$ and rapidity $\yll = 0.252$,
so the scattered protons would be within the acceptance of both AFP sides if this were an exclusive process.
However, it is found that the proton on side $A$ fails kinematic matching $|\xiAFP - \xill| < 0.005$,
so this event is likely a single-dissociative process where the side $A$ proton candidate originates from a pileup interaction.
The corresponding quantities for the highest-mass $\mu\mu$ candidate are $\mll = 319~\GeV$ and $\yll = 0.255$.
Figure~\ref{fig:paper_fig_3} illustrates detector-level distributions of dilepton acoplanarity,
mass, and rapidity after kinematic matching
with the signal samples normalized to $N_\text{obs} - N_\text{bkg}$.
 
Cross sections are measured in a fiducial region defined at particle level
with an event selection similar to that applied at detector level~\cite{fiducial-footnote}.
To reliably estimate AFP reconstruction efficiencies using tag-and-probe techniques,
the $\xiAFP$ and $\xill$ values are restricted to a tighter range $[0.035, 0.08]$
and each proton candidate is required to have an associated track in both near and far stations.
The measured cross sections are defined by
$ \sigma_\text{fid.} = \left( N_\text{obs} - N_\text{bkg} \right) / \left( \mathcal{L}\cdot C_\text{cent}\cdot C_\text{AFP} \right)$.
Here, $N_\text{obs}$ ($N_\text{bkg}$) is the number of observed data (expected background) events passing event selection,
and $C_\text{cent}$ ($C_\text{AFP}$) is an overall correction factor
accounting for the central-detector (AFP) efficiency.
The integrated luminosity, $\mathcal{L} = 14.6$~fb$^{-1}$, is measured using the LUCID-2 detector~\cite{LUCID2}
and the uncertainty is determined to be $2.4$\%~\cite{ATLAS-CONF-2019-021}.
In this tighter region, $N_\text{obs}$ is found to be \NobsEEfid~(\NobsMMfid) for the $ee$ ($\mu\mu$) channel
and $N_\text{bkg} = \NbkgEEfid~(\NbkgMMfid)$.
The event rate between the two channels differs more for the $\xi \in [0.02, 0.12]$ than $\xi \in [0.035, 0.08]$ region
because $\mu\mu$ events with low $\mll$ and high $|\yll|$ have greater selection efficiency due to trigger and reconstruction requirements.
 
The $C_\text{cent}$ factor is defined as the ratio of the number of MC events passing detector-level selection
to the number passing the particle-level fiducial requirements.
Uncertainties in $C_\text{cent}$ are estimated by varying the
electron (muon) energy (momentum) scale and resolution, and data-to-MC correction factors described in Refs.~\cite{PERF-2017-01,PERF-2015-10},
together with corrections applied to account for pileup modeling.
The dominant uncertainties for $ee$ events arise from pileup modeling (2\%) and identification (1\%),
while for $\mu\mu$ events, these correspond to pileup modeling (3\%), resolution (3\%), and scale (2\%);
other sources such as trigger and isolation efficiencies contribute 1\% or less.
Using data-driven methods described in Ref.~\cite{STDM-2016-13},
a further correction of $0.89 \pm 0.04$ is applied to $C_\text{cent}$ to account
for differences between data and simulation when modeling the luminous region
at the interaction point.
The 5\% uncertainty in this correction is evaluated as the difference between either applying this data-driven method to simulated signal samples
or imposing the $\Ntracks = 0$ requirement on these samples.
Overall, this results in $C_\text{cent}^{ee} = 0.12\pm 0.01$ $\left(C_\text{cent}^{\mu\mu} = 0.22 \pm 0.02\right)$
for the $ee$ $(\mu\mu)$ channel.
 
\begin{table}[]
\caption{\label{tab:xsec_compare}
Fiducial cross sections from the combined \textsc{herwig} and \textsc{lpair} predictions with $S_\text{surv} = 1$ and
$S_\text{surv}$ estimated using Refs.~\cite{Dyndal:2014yea,Harland-Lang:2016apc} as described in the main text.
\textsc{superchic {\scriptsize4}}~\cite{Harland-Lang:2020veo} predictions include fully kinematically dependent $S_\text{surv}$.
Uncertainties of 7\% (17\%) are assigned for predictions of the exclusive (single-dissociative) processes~\cite{theory_pred_footnote}.
The bottom row displays the measured cross sections with statistical and systematic uncertainties combined.
}
\vspace{0.3cm}
\centering
\begin{tabular}{lrr}
\toprule
$\sigma_{\textsc{herwig}+\textsc{lpair}} \times S_\text{surv}$ & $\sigma^\text{fid.}_{ee+p}$ (fb) & $\sigma^\text{fid.}_{\mu\mu+p}$ (fb)   \\
\midrule
\quad $S_\text{surv} = 1$                                                    & $15.5\pm 1.2$  & $13.5\pm 1.1$ \\
\quad $S_\text{surv}$ using Refs.~\cite{Dyndal:2014yea,Harland-Lang:2016apc} & $10.9\pm 0.8$  & $9.4\pm 0.7$  \\
\midrule
\textsc{superchic {\scriptsize4}}~\cite{Harland-Lang:2020veo}                & $12.2\pm 0.9$  & $10.4\pm 0.7$ \\
\midrule
Measurement                                                                  & $\XsecAbsSystTinyEE$ & $\XsecAbsSystTinyMM$  \\
\bottomrule
\end{tabular}
\end{table}

The $C_\text{AFP}$ factor is defined by the product $\epsilon_\text{track} \cdot \epsilon_\text{smear}$.
The track reconstruction efficiency $\epsilon_\text{track}$ is found to be $0.92 \pm 0.02$ for sides $A$ and $C$.
The near-station efficiency is estimated using a tag-and-probe method
by first selecting events with exactly one track in the far (tag) station in the acceptance common to both stations, $-12 < x_\text{AFP}<-5$~mm.
The efficiency is the fraction of these events that also have one or more tracks in the near (probe) station
satisfying $| x_\text{near} - x_\text{far}| < 2$~mm.
The tag and probe stations are inverted to measure the far-station efficiency.
It is found that $\epsilon_\text{track}$ varies with $\xiAFP$ by $2\%$, which is assigned as an additional uncertainty.
The proton resolution correction $\epsilon_\text{smear}$ is found to be $0.98\pm 0.02$ ($0.96 \pm 0.04$) for the $ee$ ($\mu\mu$) channel.
This is evaluated as the fraction of simulated signal events passing
$\xiAFP, \xill \in [0.035, 0.08]$, and $|\xiAFP - \xill|<0.005$ out of those satisfying $\xill \in [0.035, 0.08]$.
Uncertainties in $C_\text{AFP}$ are dominated by global alignment ($6\%$) evaluated by $\pm 0.3$~mm variations of $x_\text{AFP}$, and
beam optics ($5\%$) evaluated by varying the beam crossing angle by 50~$\mu$rad in the \textsc{mad-x} package.
Uncertainties involving track and cluster reconstruction are found to be less than 1\%.
The overall uncertainty in $C_\text{AFP}$ is 9\%.
 
The measured fiducial cross sections in the $ee$ and $\mu\mu$ channels are
$\sigma^\text{fid.}_{ee+p} = \XsecAbsSystShortEE$ and $\sigma^\text{fid.}_{\mu\mu+p} = \XsecAbsSystShortMM$, respectively.
Table~\ref{tab:xsec_compare} compares these with the combined \textsc{herwig} and \textsc{lpair} predictions assuming unit soft-survival factors $S_\text{surv} = 1$.
Soft-survival effects are included using an \mll-dependent reweighting of these predictions
to $S_\text{surv}$ calculated for exclusive processes from Ref.~\cite{Dyndal:2014yea};
\textsc{lpair} predictions are additionally scaled down by 15\% to account for $S_\text{surv}$ being lower for single-dissociative processes~\cite{Harland-Lang:2016apc}.
\textsc{superchic {\scriptsize4}}~\cite{Harland-Lang:2020veo} predictions include full kinematic dependence on $S_\text{surv}$ for exclusive, single-, and double-dissociative processes.
The predictions for $ee$ are higher than for $\mu\mu$ due to the looser $\eta(e)$ requirement~\cite{fiducial-footnote}.
 
In summary, forward proton scattering in association with lepton pairs produced via photon fusion, $pp \to p(\gamma\gamma\to\ellell)p^{(*)}$,
is observed with a significance exceeding $5\sigma$ in both the $ee+p$ and $\mu\mu+p$ final states
using 14.6~fb$^{-1}$ of $\sqrt{s} = 13~\TeV$ $pp$ collisions at the LHC.
These results demonstrate that the ATLAS Forward Proton spectrometer performs well in high-luminosity data taking.
Furthermore, proton tagging is introduced for cross-section measurements of photon fusion processes at the electroweak scale.


We thank CERN for the very successful operation of the LHC, as well as the
support staff from our institutions without whom ATLAS could not be
operated efficiently.
 
We acknowledge the support of ANPCyT, Argentina; YerPhI, Armenia; ARC, Australia; BMWFW and FWF, Austria; ANAS, Azerbaijan; SSTC, Belarus; CNPq and FAPESP, Brazil; NSERC, NRC and CFI, Canada; CERN; ANID, Chile; CAS, MOST and NSFC, China; COLCIENCIAS, Colombia; MSMT CR, MPO CR and VSC CR, Czech Republic; DNRF and DNSRC, Denmark; IN2P3-CNRS and CEA-DRF/IRFU, France; SRNSFG, Georgia; BMBF, HGF and MPG, Germany; GSRT, Greece; RGC and Hong Kong SAR, China; ISF and Benoziyo Center, Israel; INFN, Italy; MEXT and JSPS, Japan; CNRST, Morocco; NWO, Netherlands; RCN, Norway; MNiSW and NCN, Poland; FCT, Portugal; MNE/IFA, Romania; JINR; MES of Russia and NRC KI, Russian Federation; MESTD, Serbia; MSSR, Slovakia; ARRS and MIZ\v{S}, Slovenia; DST/NRF, South Africa; MICINN, Spain; SRC and Wallenberg Foundation, Sweden; SERI, SNSF and Cantons of Bern and Geneva, Switzerland; MOST, Taiwan; TAEK, Turkey; STFC, United Kingdom; DOE and NSF, United States of America. In addition, individual groups and members have received support from BCKDF, CANARIE, Compute Canada, CRC and IVADO, Canada; Beijing Municipal Science \& Technology Commission, China; COST, ERC, ERDF, Horizon 2020 and Marie Sk{\l}odowska-Curie Actions, European Union; Investissements d'Avenir Labex, Investissements d'Avenir Idex and ANR, France; DFG and AvH Foundation, Germany; Herakleitos, Thales and Aristeia programmes co-financed by EU-ESF and the Greek NSRF, Greece; BSF-NSF and GIF, Israel; La Caixa Banking Foundation, CERCA Programme Generalitat de Catalunya and PROMETEO and GenT Programmes Generalitat Valenciana, Spain; G\"{o}ran Gustafssons Stiftelse, Sweden; The Royal Society and Leverhulme Trust, United Kingdom.
 
The crucial computing support from all WLCG partners is acknowledged gratefully, in particular from CERN, the ATLAS Tier-1 facilities at TRIUMF (Canada), NDGF (Denmark, Norway, Sweden), CC-IN2P3 (France), KIT/GridKA (Germany), INFN-CNAF (Italy), NL-T1 (Netherlands), PIC (Spain), ASGC (Taiwan), RAL (UK) and BNL (USA), the Tier-2 facilities worldwide and large non-WLCG resource providers. Major contributors of computing resources are listed in Ref.~\cite{ATL-SOFT-PUB-2020-001}.
 
 
We are grateful to the LHC optics, collimation, machine protection and operations groups that enabled the use of the ATLAS Forward Proton spectrometer.

\printbibliography
 
\clearpage
 
\begin{flushleft}
\hypersetup{urlcolor=black}
{\Large The ATLAS Collaboration}

\bigskip

\AtlasOrcid[0000-0002-6665-4934]{G.~Aad}$^\textrm{\scriptsize 102}$,    
\AtlasOrcid[0000-0002-5888-2734]{B.~Abbott}$^\textrm{\scriptsize 128}$,    
\AtlasOrcid{D.C.~Abbott}$^\textrm{\scriptsize 103}$,    
\AtlasOrcid[0000-0002-2788-3822]{A.~Abed~Abud}$^\textrm{\scriptsize 36}$,    
\AtlasOrcid[0000-0002-1002-1652]{K.~Abeling}$^\textrm{\scriptsize 53}$,    
\AtlasOrcid[0000-0002-2987-4006]{D.K.~Abhayasinghe}$^\textrm{\scriptsize 94}$,    
\AtlasOrcid[0000-0002-8496-9294]{S.H.~Abidi}$^\textrm{\scriptsize 167}$,    
\AtlasOrcid[0000-0002-8279-9324]{O.S.~AbouZeid}$^\textrm{\scriptsize 40}$,    
\AtlasOrcid{N.L.~Abraham}$^\textrm{\scriptsize 156}$,    
\AtlasOrcid[0000-0001-5329-6640]{H.~Abramowicz}$^\textrm{\scriptsize 161}$
% ...
\bigskip
\\

% ...
$^{36}$CERN, Geneva; Switzerland.\\
% ...
$^{40}$Niels Bohr Institute, University of Copenhagen, Copenhagen; Denmark.\\
% ...
$^{53}$II. Physikalisches Institut, Georg-August-Universit\"{a}t G\"ottingen, G\"ottingen; Germany.\\
% ...
$^{94}$Department of Physics, Royal Holloway University of London, Egham; United Kingdom.\\
% ...
$^{102}$CPPM, Aix-Marseille Universit\'e, CNRS/IN2P3, Marseille; France.\\
$^{103}$Department of Physics, University of Massachusetts, Amherst MA; United States of America.\\
% ...
$^{128}$Homer L. Dodge Department of Physics and Astronomy, University of Oklahoma, Norman OK; United States of America.\\
% ...
$^{156}$Department of Physics and Astronomy, University of Sussex, Brighton; United Kingdom.\\
% ...
$^{161}$Raymond and Beverly Sackler School of Physics and Astronomy, Tel Aviv University, Tel Aviv; Israel.\\
% ...
$^{167}$Department of Physics, University of Toronto, Toronto ON; Canada.\\
% ...

\end{flushleft}

\FloatBarrier
\clearpage

\end{document}